\documentclass[french]{cfm2007}
\usepackage{amsfonts}
\usepackage[squaren]{SIunits}
\newcommand{\imag}[1]{\,\Im \mathfrak{m}\left( #1 \right)\,}
\newcommand{\real}[1]{\,\Re \mathfrak{e}\left( #1 \right)\,}
\newcommand{\w}{\omega}
\newcommand{\sgn}{\mathop{\mathrm{sign}}}
\newcommand{\derive}[1]{\frac{d {#1}}{dt}}
\newcommand{\deriven}[2]{\frac{d^{#1} #2}{dt^{#1}}}

\begin{document}
\cfmtitre{Seuils d'instabilité d'un instrument de musique à anche simple :\\ approche modale}
\cfmauteur{Fabrice Silva \& Jean Kergomard}
\cfmadresse{Laboratoire de Mécanique et d'Acoustique, UPR CNRS 7051\\
    31 chemin Joseph Aiguier, 13402 Marseille cedex 20\\
    \{silva,kergomard\}@lma.cnrs-mrs.fr}

\vspace{-1mm}
\cfmresume{De nombreux instruments de musique (comme les instruments à vent ou à cordes frottées) ont un fonctionnement reposant sur l'entretien d'oscillations par une action continue du musicien. Dans le cas de la clarinette, l'anche, vibrant au gré des ondes acoustiques dans le corps de l'instrument, module l'apport d'énergie qu'est le souffle du musicien. Nous nous intéressons au seuil d'oscillation, valeur de la pression dans la bouche à partir de laquelle un son peut prendre naissance et cherchons à comprendre comment le comportement dynamique de l'anche, par sa collaboration avec les résonances du tuyau, influe sur ce seuil. Nous présentons ici une méthode reposant sur la décomposition modale du résonateur acoustique qui autorise une étude au-delà du seuil, et la confrontons avec une résolution directe du problème de recherche de seuil.}
\vspace{-1mm}
\cfmabstract{Many musical instruments, as for example woodwind instruments, flute or violins, are self-sustained oscillating systems, i.e. musician enacts as a continuous energy source to drive an oscillation in the passive resonator, the body of the instrument, by means of a nonlinear coupling. For single reed instruments like clarinet, there exists a minimal value of mouth pressure beyond which sound can appear. This paper deals with the analysis of this oscillation threshold, calculated using a modal decomposition of the resonator, in order to have a better comprehension of how reed characteristics, such as its strength and its damping, may influence the attack transient of notes.}
\vspace{-1mm}
\cfmkeywords{acoustique musicale, décomposition modale, seuil d'oscillation}

\section{Introduction}
\vspace{-1mm}
\par Dans les instruments de musique à anche simple tels que la clarinette comme dans d'autres systèmes comprenant un guide d'onde muni d'une valve libre à une de ses extrémités, des oscillations peuvent apparaître dès lors que la pression d'alimentation dépasse une valeur critique. Ce seuil d'oscillation, directement lié à la notion de facilité d'émission du son pour les instruments de musique, a fait l'objet d'un grand nombre d'études, aussi bien expérimentales (cf. \cite{DFG+06}, \cite{WB74}) qu'analytiques (cf. \cite{DGK+}, \cite{Gra94}) sur des modèles simplifiés, comme par exemple le modèle dit de Raman.
 La motivation principale de ce travail est de proposer une méthode numérique d'investigation des seuils d'oscillation reposant sur l'étude linéaire de stabilité. L'originalité de l'approche consiste en une description modale du résonateur acoustique, généralisant ainsi des travaux précédents portant sur l'interaction entre un mode d'anche lippale et un mode de tuyau dans le cas des cuivres, cf. \cite{CGC00}, l'anche étant ici décrite par son premier mode de flexion. La méthode de décomposition modale a été précédemment utilisée par \cite{Deb04} pour le calcul des auto-oscillations dans le domaine temporel. L'intérêt de la méthode est de pouvoir aisément être transposée aux différents instruments à anche, aussi bien pour les instruments coniques que pour les anches doubles et les cuivres.

\section{Le modèle linéarisé}
\par Nous modélisons la clarinette comme le couplage d'un résonateur acoustique (le corps de l'instrument) et d'un élément vibrant (l'anche) par l'écoulement d'air à travers le canal d'anche sous l'effet de la différence de pression entre la bouche du musicien et l'entrée du tuyau.
\par Un des objectifs du travail est de caractériser la manière dont le comportement dynamique de l'anche intervient dans le seuil d'oscillation du système couplé. Nous limitons ici cette dynamique au premier mode de flexion de l'anche, le mouvement de l'anche pouvant ainsi être décrit par une équation différentielle du 2\degre  ordre liant le déplacement de son extrémité libre $y(t)$ (par rapport à sa position au repos $y_0$) à la différence entre la pression dans la bouche $p_m$ et dans le bec $p(t)$~:
\begin{equation}
\frac{1}{\w_r^2} \deriven{2}{y} +\frac{q_r}{\w_r} \derive{y} +\left( y(t)-y_0\right)
 = \frac{1}{K}\left( p(t)-p_m \right)
\end{equation}
où $K$ est un coefficient surfacique de raideur qui peut être évalué à partir d'un essai statique, $\w_r$ et $q_r$ étant la pulsation propre et le coefficient d'amortissement de l'anche estimés à partir d'un essai de laché (\cite{WB74}).
\par Le couplage entre vibration de l'anche et ondes acoustiques dans le corps de l'instrument est réalisé de manière aéro-acoustique par l'écoulement de l'air de la bouche vers le bec de la clarinette à travers l'ouverture entre l'anche et la table. L'utilisation du théorème de Bernoulli a été justifiée par \cite{Hir} sous les hypothèses d'écoulement stationnaire et incompressible (masse volumique $\rho$), de grande différence de section entre la bouche et le canal et de dissipation de l'énergie cinétique du jet par turbulence à la sortie du canal. En supposant de plus l'ouverture rectangulaire de largeur $W$ et de hauteur $y(t)$, il permet d'établir la relation non-linéaire entre le débit $u(t)$ entrant dans l'instrument et la différence de pression~:
\begin{equation}
u(t) = Wy(t) \sqrt{\frac{2}{\rho}} \sgn{\left(p_m-p(t)\right)}\sqrt{|p_m-p(t)|}
\quad \mbox{si $y(t)\geq 0$, sinon $u(t)=0$.}
\end{equation}
L'étude se portant au voisinage du seuil d'oscillation afin de déterminer quels sont les modes du système couplé qui naissent en premier, nous nous plaçons dans le cadre d'une étude linéaire de stabilité, ce qui autorise la linéarisation de la caractéristique de couplage et le passage dans le domaine fréquentiel (avec une convention de dépendance temporelle en $\exp{j\w t}$) pour de faibles écarts au régime statique. Il vient alors~:
\begin{equation}
U(\w) = \frac{2p_m}{\rho}W Y(\w)
    -\frac{Wy_0}{\rho}\left(1-\frac{p_m}{Ky_0}\right) P(\w)
\,\mbox{ et }\,
Y(\w) = \frac{1}{K}\frac{1}{1+jq_r\frac{\w}{\w_r}-\frac{\w^2}{\w_r^2}}P(\w)
\end{equation}
L'admittance aéro-acoustique (réduite) vue par le résonateur acoustique est donc~:
\begin{equation}
\mathcal{Y}_a(\w)=\frac{Z_cu_1(\w)}{p_1(\w)}
=\zeta \sqrt{\gamma}\left(D(j\w)-\frac{1-\gamma}{2\gamma}\right)
\mbox{ avec } D(j\w)=\frac{1}{1+jq_r\frac{\w}{\w_r}-\frac{\w^2}{\w_r^2}}
\end{equation}
où $Z_c=\rho c/S$ est l'impédance caractéristique du tuyau (liée à sa section $S$ et à la célérité $c$ du son en espace libre), $\gamma$ et $\zeta$ sont deux paramètres adimensionnés représentant la pression dans la bouche et la \og pince\fg du musicien~:
\begin{equation}
\gamma=\frac{p_m}{Ky_0}\quad \mbox{et}\quad
\zeta=Z_c W\sqrt{\frac{2y_0}{\rho K}}
\end{equation}
\par Le résonateur acoustique est décrit par son impédance d'entrée $\mathcal{Z}_e$, en faisant l'hypothèse d'un tuyau cylindrique de longueur $L$ et de rayon $r$ ouvert à son extrémité (l'impédance de rayonnement est considérée nulle), et en prenant en compte un amortissement des ondes $\alpha$ par des phénomènes visco-thermiques aux parois. Cette impédance peut être décomposée sur les $N$ premiers modes propres du résonateur passif définis par les pulsations propres $\Omega_n$ (ce sont celles du tuyau sans pertes légèrement abaissées du fait de la dispersion résultant des pertes visco-thermiques) et leurs coefficients d'amortissement $\alpha_n$ (estimées à partir de la valeur de $\alpha$ à la pulsation propre associée)~:
\begin{equation}
\mathcal{Z}_e(\w)=\frac{p_1(\w)}{Z_c u_1(\w)}=j\tan{\left(\frac{\w}{c}-j\alpha·\right)L}
\quad \simeq \quad
\mathcal{Z}_N(\w)\simeq \frac{2c}{L}
\sum_{n=1}^N \frac{j\w}{\Omega_n^2+2j\w\alpha_n c+(j\w)^2}
\label{eq:Zmodal}
\end{equation}
$c$ étant la célérité du son en espace libre.

\section{Modes propres du système linéarisé et seuils d'oscillation}
L'équation caractéristique du système linéarisé est obtenue en égalant impédance aéro-acoustique et impédance d'entrée du résonateur. Avec la formulation modale de l'impédance du tuyau, il vient~:
\begin{equation}
\frac{2c}{L}\sum_{n=1}^N \frac{j\w}{\Omega_n^2+2j\w\alpha_n c+(j\w)^2}
=\frac{1}{\zeta\sqrt{\gamma}}\left(D(j\w)-\frac{1-\gamma}{2\gamma}\right)^{-1}.
\label{eq:caract}
\end{equation}
\par La démarche proposée est celle de la recherche du seuil d'oscillation par les méthodes d'étude de stabilité linéaire. La recherche des solutions $j\w$ de l'équation caractéristique pour une configuration donnée ($L$, $r$, $\w_r$, $q_r$, $\gamma$ et $\zeta$ fixés) permet d'obtenir les modes propres du système linéarisé autour du régime statique. Les parties imaginaire et réelle correspondent respectivement à la fréquence et à l'amortissement des différents modes propres~:
\begin{equation}
e^{j\w t} = e^{\real{j\w}t} e^{j\imag{j\w}t}
\end{equation}
\par Un des avantages de la formulation modale de l'impédance d'entrée du résonateur est de pouvoir faire apparaître l'équation caractéristique sous forme d'un polynôme en $j\w$, ce qui permet l'utilisation d'algorithmes efficaces pour obtenir les $(2N+2)$ solutions réelles ou complexes conjuguées. La question du nombre de modes considérés, c'est-à-dire de l'ordre de troncature du développement modal de la formulation (\ref{eq:Zmodal}), est examinée plus bas.
\par
\begin{figure}[hbt]
    \centering
    \includegraphics*[]{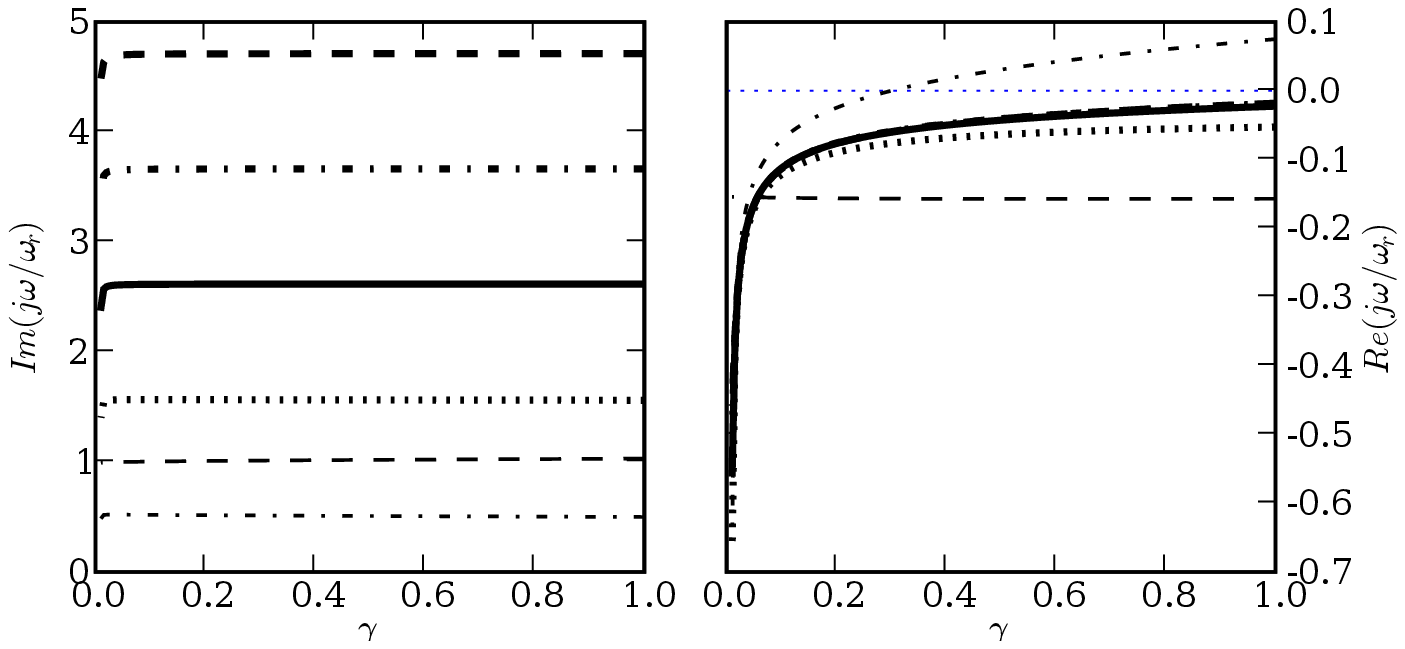}
    \caption{\label{fig:Fig2}\'Evolution des fréquences propres complexes en fonction de la pression dans la bouche~:\newline $L=16\centi\metre$, $r=7\milli\metre$, $\w_r=2\pi\times 1000\radian\per\second$, $q_r=0.3$, $\zeta=0.2$.}
\end{figure}

 Il est possible d'observer l'évolution des fréquences propres complexes en fonction de $\gamma$, image de la pression dans la bouche, la recherche des modes propres étant répétée pour chaque valeur de $\gamma$. Sur la figure~\ref{fig:Fig2}, on peut visualiser une première plage de pression où toutes les solutions sont amorties ($\real{(j\w)}<0$) : le régime statique est le seul régime stable pour de faible pression d'alimentation. Il existe ensuite une valeur de $\gamma$ pour laquelle une unique solution a une partie réelle nulle, les autres étant à partie réelle négative : c'est le seuil d'instabilité du régime statique. Pour des valeurs supérieures, il existe au moins un mode propre amplifié : le régime statique est instable. Dans ce dernier cas, les oscillations peuvent alors croître exponentiellement jusqu'à ce que la non-linéarité contenue dans la caractéristique de débit vienne saturer l'amplitude des oscillations.
\begin{figure}[hbt]
    \centering
    \includegraphics*[]{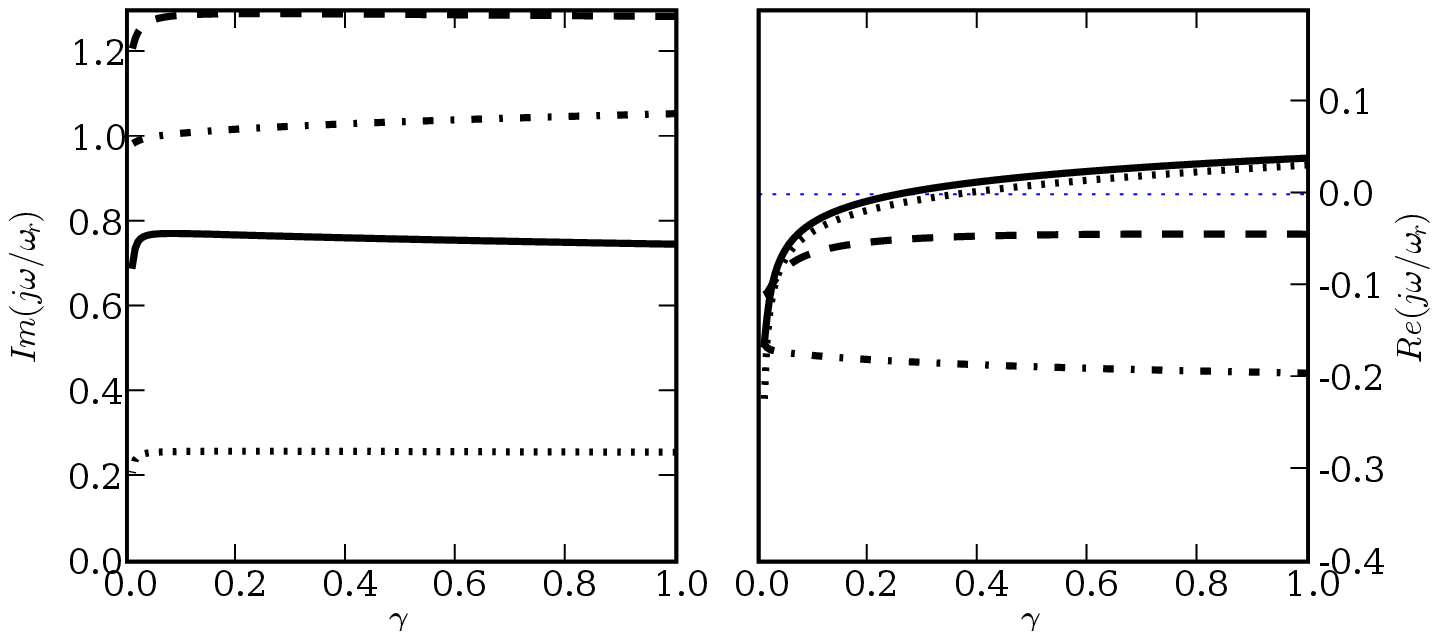}
    \caption{\label{fig:Fig3}\'Evolution des fréquences propres complexes en fonction de la pression dans la bouche~:\newline $L=32\centi\metre$, $r=7\milli\metre$, $\w_r=2\pi\times 1000\radian\per\second$, $q_r=0.3$, $\zeta=0.2$.}
\end{figure}
\par Sur la figure~\ref{fig:Fig2} est représentée l'évolution des modes propres dans le cas où $\w_rL/c=3$. Les fréquences propres du système couplé ont des valeurs proches des fréquences de résonance du tuyau isolé ($\w \simeq (n-1/2)\pi c/L$) et de l'anche seule ($\w=1$), les fréquences propres variant faiblement avec la pression d'alimentation. Le seuil d'instabilité du régime statique est de $\gamma\simeq 0.3$, avec une oscillation émergente à une fréquence correspondant à la première résonance du tuyau. Dans le second cas (figure~\ref{fig:Fig3}, $\w_rL/c=6$), le seuil, obtenu pour une valeur de pression proche de la précédente, est associé à la seconde résonance du tube, celle-ci correspondant à une fréquence légèrement inférieure à la résonance d'anche. L'hypothèse d'un comportement de l'anche réduit à sa raideur aboutirait à un seuil associé au premier mode du tuyau inférieur aux seuils des modes supérieurs du fait de l'augmentation des pertes avec la fréquence, cependant l'interaction entre les résonances acoustiques avec la résonance mécanique est responsable de cette inversion de seuils entre premier et second mode, comme expliqué par \cite{WB74}.
\par L'ordre de troncature de la décomposition modale a été choisi selon un critère lié à la résolution de notre algorithme de recherche de seuil et à la vitesse de convergence du seuil $\gamma$ avec le nombre de modes pris en compte. La troncature à $N=20$ modes a été retenue pour les simulations numériques, elle correspond à une modification du résonateur engendrant une déviation du seuil d'oscillation de moins de $0.01$, qui est la tolérance utilisée lors de la recherche de zéros.
\subsection*{Comparaison avec une résolution directe}
\par La détermination du seuil d'oscillation avec l'approche modale a été réalisée numériquement avec la méthode de la position fausse en s'intéressant à la fonction~:
\begin{equation}
F:\gamma\longrightarrow \max{\left\{
    \real{j\w}\quad/\quad j\w \mbox{ solution de (\ref{eq:caract})}
    \right\}},
\end{equation}
le changement de signe de cette fonction correspondant à la perte de stabilité du régime statique.
\par Nous avons confronté (cf. Table~\ref{table}) les résultats obtenus par cette méthode avec ceux issus d'une résolution directe de l'équation caractéristique en $(\theta,\gamma)$ avec $\theta=\w/\w_r$ à valeur réelle. Une des difficultés de la résolution directe est liée à la nature transcendentale de l'équation caractéristique avec la formulation de l'impédance d'entrée en $\tan{}$. Le nombre de solutions étant infini, il faut s'assurer que la recherche des solutions amène bien au seuil le plus bas.
\begin{table}[hbt]
    \begin{center}
    \begin{tabular}[htp]{|c||cc|c||cc|c|}
        \hline
        $k_rL$ & $\theta_{DM}$ & $\theta_{RD}$ & $\Delta \theta$
            & $\gamma_{DM}$ & $\gamma_{RD}$ & $\Delta \gamma$\tabularnewline
        \hline\hline
        8.5 & 0.185 & 0.186 & 0.5\%     & 0.43 & 0.43 & 0.1\%    \tabularnewline
        2 & 0.747 & 0.747 & 0.1\%         & 0.30 & 0.30 & 0.6\%    \tabularnewline
        1 & 1.022 & 1.024 & 0.2\%       & 3.86 & 3.82 & 1\%        \tabularnewline
        0.81 & 1.033 & 1.034 & 0.1\%    & 8.77 & 8.71 & 0.7\%    \tabularnewline
        \hline
    \end{tabular}
    \caption{\label{table} Comparaison des seuils obtenus par l'approche modale (notée DM) et par résolution directe (RD) pour $r=7\milli\metre$, $\w_r=2\pi\times 750\radian\per\second$, $q_r=0.4$, $\zeta=0.13$.}
    \end{center}
\end{table}
\section{Couplage d'un mode d'anche et d'un mode du tuyau}
Dans ce cas simple, on obtient des résultats analytiques sur la manière dont l'écoulement va modifier les modes des systèmes. Avec l'hypothèse d'un tuyau à un seul mode et en ne tenant compte que des fréquences positives, une approximation de $D$ et $\mathcal{Z}_1$ est donnée par~:
\begin{equation}
D^{-1}(\w)\simeq -\frac{2}{\w_r}(\w-\w_r-jq_r\w_r/2) \quad\mbox{et}\quad
\mathcal{Z}_1^{-1}(\w) \simeq j\frac{L}{c}(\w-\Omega_n-j\alpha_n c),
\end{equation}
ce qui, à partir de l'équation~(\ref{eq:caract}), aboutit à la relation de couplage~:
\begin{equation}
(\w-\w_r^+)(\w-\w_n^+)\simeq j\frac{\w_r c}{2L}\zeta\sqrt{\gamma}
\,\mbox{ avec }\,
\w_r^+=\w_r\left(1+j\frac{q_r}{2}\right)
\,\mbox{ et }\,
\w_n^+=\Omega_n+j\alpha_n c+j\zeta\frac{c}{2L}\frac{1-\gamma}{\sqrt{\gamma}}.
\label{eq:2modes}
\end{equation}
On retrouve le fait que les deux modes sont couplés par l'écoulement, mais le mode acoustique à considérer n'est pas celui du tuyau \og ouvert-fermé \fg : puisque $\gamma<1$ (sinon l'anche plaque en régime statique), les pertes sont augmentées, ce qui s'explique par le comportement acoustique résistif de l'écoulement à l'entrée du tuyau.
\par Un résultat remarquable que l'on obtient sur ce modèle simplifié est que lorqu'une solution devient instable (l'exponentielle temporelle associée devient croissante), l'autre solution est toujours stable. En effet, la somme des deux solutions vaut $(\w_r^+ + \w_n^+)$, et cette quantité est toujours à partie imaginaire positive. Dans cette configuration à deux modes, les seuils étudiés sont donc bien des seuils d'instabilité du régime statique.

\section{Conclusions}
\par La comparaison avec une méthode de détermination directe des seuils d'instabilité du régime statique permet de valider la démarche proposée dans cette communication. Il est possible d'expliquer les faibles écarts constatés pour $\gamma$ par le caractère itératif de la méthode utilisée, avec une condition d'arrêt portant sur un encadrement de la solution à $\delta\gamma = 0.01$. Un autre facteur est la troncature de la décomposition modale de l'impédance d'entrée du résonateur. En effet, la troncature correspond à une modification du guide d'onde, d'autant moins significative que le nombre de modes pris en compte est grand. Toutefois, l'écart entre les résultats des deux méthodes est suffisamment faible pour ne pas chercher à travailler avec un plus grand nombre de modes.
\par Un des intérêts majeurs de cette démarche est de pouvoir être appliquée pour tout résonateur dont l'impédance d'entrée peut se mettre sous forme modale, puisqu'il suffit de connaître l'amplitude, la pulsation et le facteur de qualité de chacun des modes, ce qui peut fait par ajustement de paramètres sur une mesure fréquentielle d'impédance d'entrée. Il est également possible d'étendre à des éléments mécaniques à plusieurs modes de flexion et/ou de torsion, à la prise en compte de phénomènes supplémentaires comme le débit d'anche.
\par Outre l'étude des seuils d'instabilité du régime statique, l'approche proposée permet également de s'intéresser à ce qu'il se passe au-delà du seuil. En effet, dans la limite de validité de la linéarisation de la caractéristique de débit, le modèle linéarisé reste adapté à l'étude des démarrages des auto-oscillations, notamment en ce qui concerne la vitesse des transitoires, tout au moins jusqu'à ce que les non-linéarités viennent saturer l'amplitude des ondes acoustiques.

\medskip
\par Les travaux présentés dans cette communication ont été réalisés dans le cadre de CONSONNES, projet financé par l'Agence Nationale de la Recherche.
\bibliographystyle{CFM07}
\bibliography{BiblioGraphie}
\end{document}